\documentclass[aps,pre,showpacs,showkeys, twocolumn,superscriptaddress,
floats,floatfix,preprintnumbers,longbibliography,multicol]{revtex4-2}
\usepackage{amssymb,amsmath}
\usepackage{bm,url,graphics,graphicx,subfigure,epsfig,rotating} 
\usepackage{mathrsfs} 
\usepackage{color,soul}
\graphicspath{{./}{Fig/}}
\usepackage{braket}
\usepackage{hyperref}
\usepackage{multirow} 
\usepackage{booktabs}
\usepackage{makecell}
\usepackage[normalem]{ulem}   
\usepackage[dvipsnames, table, xcdraw]{xcolor}
\usepackage{relsize}
\usepackage{euler}
\usepackage{tikz}
\usepackage{feynmp-auto}
\usepackage{CJKutf8}
\def \dc  {d_{\rm c}}

\def \gastn {g^{\ast}_{\rm 0}}
\def \gasto {g^{\ast}_{\rm 1}}
\def \Mo  {M_{\rm 1}}
\def \nuo  {\nu_{\rm 1}}
\def \Do  {D_{\rm 1}}
\def \kkp {\bm{k}_{+}}
\def \kkm {\bm{k}_{-}}
\def \kp {k_{+}}
\def \km {k_{-}}

\def \rr {\bm{r}}
\def \fh {\hat{f}}
\def \Mn {M_{\rm o}}
\def \nun {\nu_{\rm o}}
\def \Dn {D_{\rm o}}
\def \kko {\bm{k}_{\rm 1}}
\def \kkt {\bm{k}_{\rm 2}}
\def \kkth {\bm{k}_{\rm 3}}
\def \kkf {\bm{k}_{\rm 4}}
\def \oo {\omega_{\rm 1}}
\def \ot {\omega_{\rm 2}}
\def \oth {\omega_{\rm 3}}
\def \of {\omega_{\rm 4}}
\def \kvo {\vec{k}_{\rm 1}}
\def \kvt {\vec{k}_{\rm 2}}
\def \kvth {\vec{k}_{\rm 3}}
\def \kvf {\vec{k}_{\rm 4}}

\def \SI   {S^{\rm I}}

\def \intg {\int^{>}}
\def \Srep {S_{\rm replica}}
\def \phig {\phi^{>}}
\def \Phig {\Phi^{>}}
\def \phil {\phi^{<}}
\def \Phil {\Phi^{<}}

\def \Phia {\Phi_{\alpha}}
\def \Phib {\Phi_{\beta}}
\def \phia {\phi_{\alpha}}
\def \phib {\phi_{\beta}}

\def \dab {\delta_{\alpha\beta}}

\def \dl {\delta \ell}

\def \Kd {K_{\rm d}}

\def \Sd   {S_{\rm d}}

\def \Vd   {V_{\rm d}}

\def \An  {A_{\rm n}}
\def \Bn  {B_{\rm n}}
\def \Ao  {A_{\rm 1}}
\def \Bo  {B_{\rm 1}}

\def \half {\frac{1}{2}}

\def \mZ    {\mathcal{Z}}

\def \mD {\mathcal{D}}

\def \mm  {\bm{m}}
\def \mD   {\mathcal{D}}
\def \mM   {\mathcal{M}}

\def \qq     {\bm{q}}

\def \pp  {\bm{p}}
\def \mm  {\bm{m}}

\def \mL   {\mathcal{L}}

\def \mP {\mathcal{P}}

\def \kk  {{\bm k}}
\def \kv  {\vec{k}}

\def  \xx  {{\bm x}}
\def \yy {\bm{y}}

\def \grad {{\bm \nabla}}

\def \lap {\nabla^2}
\def \delt {\partial_t}
\def \Dt {\tilde{D}}

\def \Dot {\textcolor{red}{\bullet}}

\newcommand{\avg}[1]{\left\langle #1\right\rangle}
\newcommand{\avgm}[1]{\left\langle #1\right\rangle_{\rm m}}

\def\i{{\rm i}}

\newcommand{\eq}[1]{~(\ref{#1})}

\newcommand{\Fig}[1]{Fig.~\ref{#1}}

\newcommand{\bfig}{\begin{figure}}
\newcommand{\efig}{\end{figure}}
\newcommand{\bc}{\begin{center}}
\newcommand{\ec}{\end{center}}
\newcommand{\bea}{\begin{eqnarray}}
\newcommand{\eea}{\end{eqnarray}}

\def \eee     {\bm{e}}

\def \ec     {\eee_{\rm 4}}

\def \hot {\text{h.o.t.}}

\begin{document}

\title{Renormalization group analysis of noisy neural field}

\author{Jie Zang\begin{CJK}{UTF8}{gbsn} (臧杰)\end{CJK}}
\affiliation{School of Mathematics, South China University of Technology,
  Guangdong, China.}
\affiliation{Division of Computational Science and Technology,
  School of Electrical Engineering and Computer Science,
  KTH Royal Institute of Technology, Stockholm, Sweden}
\author{Pascal Helson}
\affiliation{Division of Computational Science and Technology,
  School of Electrical Engineering and Computer Science,
  KTH Royal Institute of Technology, Stockholm, Sweden.
  Science for Life Laboratory, Sweden.}
\author{Shenquan Liu\begin{CJK}{UTF8}{gbsn} (刘深泉)\end{CJK}}
\affiliation{School of Mathematics, South China University of Technology,
  Guangdong, China.}
 \author{Arvind Kumar}
 \email{arvkumar@kth.se}
\affiliation{Division of Computational Science and Technology,
  School of Electrical Engineering and Computer Science,
  KTH Royal Institute of Technology, Stockholm, Sweden.
  Science for Life Laboratory, Sweden.}
\author{Dhrubaditya Mitra}
\email{dhruba.mitra@gmail.com}
\affiliation{Nordita, KTH Royal Institute of Technology and
Stockholm University, Hannes Alfv\'ens v\"ag 12, 10691 Stockholm, Sweden}
\date{\today}

\begin{abstract}
Neurons in the brain show great diversity in their individual properties
and their connections to other neurons.
To develop an understanding of how neuronal diversity contributes to
brain dynamics and function at large scales
we start with a linearized version of the Wilson-Kowan model
and introduce a random anisotropy to inter-neuron connection.
The resultant model is  Edwards-Wilkinson model with
a random anisotropic term. 
Averaging over the quenched randomness with the replica method 
we obtain a bi-quadratic nonlinearity.
We use Wilsonian dynamic renormalization group to analyze this
model. 
We find that, up to one loop order, for dimensions higher than two,
the effect of the noise is to change dynamic exponent
from two to one.
\end{abstract}

\maketitle

\section{Introduction}
Neurons in the brain show great diversity in their
shapes~\cite{peng2021morphological}, biophysical
properties~\cite{lim2018development} and spiking
patterns~\cite{le2024prefrontal,insanally2024contributions}.
This diversity not only can be observed across brain regions but also within
a small network of neurons within a brain region~\cite{stanley2020continuous,
  cembrowski2018continuous,isbister2023modeling}.
Even neurons within a single ``cell type'' does not have the same structure
and response~\citep{angelo2012biophysical, scala2021phenotypic,
  cembrowski2018continuous}.
How neuronal diversity contributes to brain dynamics and function is an
important question in modern neuroscience.
Usually this question is studied using highly simplified models of cortical
connectivity such as random networks~\cite{van1996chaos,
  brunel2000dynamics, kumar2008high, el2009master} with fixed
distance-independent connectivity or locally connected random
networks~\cite{ermentrout1998neural, spreizer2019space, rosenbaum2017spatial}.
Interesting suggestions have been made on how neuron diversity renders the 
brain networks robust \cite{rich2022loss,hutt2023intrinsic}, improves stimulus
encoding \cite{insanally2024contributions} and contributes to computational
repertoire of the network~\cite{marsat2010neural, gast2024neural}.
But the effect of neuron properties on network activity/function is contingent
on the network activity regime~\cite{sahasranamam2016dynamical}.

However, neurons in the brain are not wired randomly and their connectivity
is constrained by both their physical shapes~\cite{jiang2015principles} and
chemical signatures~\cite{barabasi2020genetic}.
To a reasonable approximation, we can assume that connectivity decreases with
distance in a monotonic fashion~\cite{xu2016primary}.
Dynamics of such spatial networks with homogeneous neuron properties and
connectivity have been extensively studied~\cite{ermentrout1998neural,
  coombes2006neural, bressloff2001geometric, rosenbaum2017spatial,
  spreizer2017activity, tiberi2022gell}.
In spatial networks, neuron diversity is introduced by choosing neuron
properties from a distribution~\cite{kumar2008conditions} and typically
spatial correlations in neuron diversity are set to zero.
In terms of connectivity, in spatial network models all neurons are assumed to
have the same connectivity kernel and andy diversity in connection arises
simply due to finite size effects.
In these networks, spatial correlation arise in the neuronal connectivity as
the connectivity kernel of neighboring neurons overlaps, therefore,
similarity in the connectivity also decays
monotonically~\cite{mehring2003activity}.
However, unlike in random networks, in spatial networks it is important to
consider how heterogeneity in both connectivity and neuron properties are
spatially correlated.
Spreizer et al.~\cite{spreizer2019space} showed that when neural connectivity
is asymmetric (i.e. neurons make some connections preferentially in a certain
direction) and the preferred connection of neighboring neurons are similar,
travelling waves and spatio-temporal sequence can arise depending on how the
spatial correlation decays as a function of distance.
Similar effects are likely when spatial correlations are introduced in neuron
properties.
Besides these insights, dynamical consequences of the spatial distribution of
neuron and connectivity diversity are poorly understood. 

Here we introduce a theoretical framework to understand and identify
under which conditions spatial correlations in properties of neurons and their
connectivity may affect network dynamics and give rise to non-trivial
activity patterns.

We assume that the  properties and connectivity of the neurons change at a
much slower time scales than the network activity dynamics.
Therefore, we can consider heterogeneity in neuronal properties and
connectivity as quenched noise.
We use the framework of classical stochastic
fields~\cite{buice2007field, buice2010systematic} to develop a theory of
fluctuating activity in excitatory-inhibitory (EI) networks.
Recently, Tiberi et al. \cite{tiberi2022gell} have applied the
dynamic renormalization group (RG) technque to a prototypical neural field
model, specifically a simplified version of the Wilson-Cowan
model~\cite{wilson1972excitatory, wilson1973mathematical,coombes2006neural}.
The Wilson-Cowan model is a nonlinear integro-differential equation with
constant coefficients driven by a stochastic noise.
Under certain simplifying assumptions Tiberi et al. \cite{tiberi2022gell}
reduced this to a nonlinear partial differential equation (PDE) with constant
coefficients driven by a stochastic noise.
We introduce quenched randomness into this model
adding a random anisotropic term.
making the model anisotropic at small scales,
but \textit{statistically} homogeneous and isotropic at large scales.
We then average over the quenched noise using the
replica trick and extract an effective theory at large scales. 
We then analyze the effective theory using dynamic RG.
Our calculations show that, up to one loop order, in dimensions
strictly greater than two, the role of the noise is to
change the dynamic exponent from two to one, i.e.,
noise changes diffusive dynamic to advective. 
Thus, our work shows the effect of  diversity in neuronal network
can generate novel emerging dynamical states. 
\section{Model}
We start with a neural field following the stochastic Wilson-Cowan equation:
\begin{equation}\label{eq:Wilson-cowan}
    \tau \frac{d \phi}{d t}=-l(\phi)+ w * f(\phi)+\sqrt{\tau} I,
\end{equation}
where $\phi(\xx, t)$ represents the neural activity evolving in time $t$
over the spatial domain $\xx \in \mathbb{R}^d$. 
Here, $\tau$ is the characteristic timescale, $w(\xx-\yy)$ is the connectivity
kernel which weights the input from the neural state at position $\xx$ to that
at position $\yy$ through spatial convolution denoted by $*$, and $I$ is a
Gaussian noise with zero mean and correlation
\begin{equation}
  \avg{I(\xx,t)I(\yy,s)} = D\delta^d(\xx-\yy)\delta(t-s)\/.
  \label{eq:noise}
  \end{equation}
Starting from Wilson-Cowan model, \citet{tiberi2022gell} derived a neural
field under the assumptions of homogeneity and isotropy:
\begin{equation}
    \delt \phi =\sum_{n=1}^{\infty}\left[-\An+ \Bn\lap\right] \phi^n + I\/.
\label{eq:WC}
\end{equation}
Here $\An$, $\Bn$ are constants and $\lap$ is the Laplacian operator
accounting for spatial diffusion.
This model is derived from the  Wilson-Cowan equation by expanding
the functions $l(\phi)$ and $f(\phi)$ in a Taylor series in $\phi$
and by expanding the kernel and then ignoring spatial derivatives
or fourth order and higher and setting the characteristic
time scale $\tau=1$. 
In the spirit of constructing Landau--like field theories,
\eq{eq:WC} is the simplest model we can write for a scalar field ($\phi$)
under the given symmetries (homogeneity and isotropy) and
two additional constrains: the interactions between neurons are local
and we ignore derivatives of fourth order and higher. 
\subsection{Scaling}
Let us first study \eq{eq:WC} under the rescaling of space and time. 
We first select a subset of terms of \eq{eq:WC} such that
\begin{equation}
  \delt \phi = \Bo\lap\phi + I.
\label{eq:EW}
\end{equation}
This is the well-known Edwards-Wilkinson (EW)
equation~\citep{chui1978dynamics, edwards1982surface},
see also Ref.~\citep[][Chapter 5]{Barabasi1995fractal} for a
pedagogic introduction. 

The EW equation under rescaling
$x \to bx$, $t\to b^{z}t$, $\phi \to b^{\alpha}\phi$,
and $I \to b^{\lambda}I$ gives
\begin{subequations}
\begin{align}
  b^{2\lambda} &= b^{-(d+z)} \\
  \text{and}\quad \delt \phi &= b^{z-2}\Bo\lap\phi + b^{z+\lambda-\alpha}I.
\end{align}
\end{subequations}
The first equation follows from the scaling of the noise in \eq{eq:noise}.
Demanding that the EW equation remains invariant under rescaling
we obtain
\begin{equation}
  z = 2, \quad \lambda = -\frac{z+d}{2}\quad\text{and}\quad
  \alpha = \frac{2-d}{2}\/,
  \label{eq:EWs}
\end{equation}
which gives the well--known EW
exponents~\citep[][Eq. 5.16]{Barabasi1995fractal}.
Next we note how \eq{eq:WC} behaves under the same rescaling,
\begin{multline}
  \label{eq:WCs}
    \delt \phi = (-b^{z}\Ao + \Bo\lap)\phi + I \\
    +\sum_{n=2}^{\infty}\left[-b^{(n-1)\alpha+z}\An+ 
        b^{(n-1)\alpha+z-2}\Bn\lap\right] \phi^n \/.
\end{multline}
Substituting the values of $\alpha$ and $z$ from \eq{eq:EWs} we find that
in the limit $b\to \infty$ (coarse graining):
\begin{itemize}
\item The terms $-\An\phi^n$ scale as $b^{(n-1)(2-d)/2 +2}$.
  If the  exponent $(n-1)(2-d)/2 +2 < 0$  then all these terms
  go to zero. This happens for $d > 2 + 4/(n-1)$.
  At $d=2$ all these terms diverge as $b^2$.
\item The terms $-\Bn\phi^n$ for $n\ge 2$ scale as  
  $b^{(n-1)(2-d)/2}$. If the exponent $(n-1)(2-d)/2 < 0$
  then all these terms go to zero. This happens for $d>2$.
\end{itemize}
Thus we conclude that at $d = 2$ \eq{eq:WC} must have the following
form under coarse--graining
\begin{equation}
  \delt \phi = 
  \sum_{n=1}^{\infty}\left[-b^{2}\An+ \Bn\lap\right]\phi^n + I\/.
\label{eq:WCc}
\end{equation}
We are left with two choices. One, we must have the infinity of terms
$\An\phi^n$ and $\Bn\lap\phi^n$ present in the model. 
The simple scaling of EW model does not hold any longer.
Two, we must ignore -- by hand set $\An=0$ -- and still have the
infinity of terms $\Bn\lap\phi^n$.
In this case too, the simple EW scaling does not hold because
there is no \textit{a-priori} reason why the terms nonlinear in $\phi$
should obey the simple scaling obtained from the linear EW equation.
\citet{tiberi2022gell} have taken the second choice
and then arbitrarily limited $\Bn$ up to $n=2$. 
They stated that due to the balance between excitatory and inhibitory
inputs in brain networks the terms $\An\phi^n$ must be zero.
Fundamentally speaking, this is a mistaken conclusion.
Even a very small $\An$ for all dimensions $d \ge 2$ will coarse--grain to
very large values.
In what follows, we do not address this problem. 
\subsection{Model with quenched noise}
The Wilson-Cowan model applies at a scale that contains many neurons
-- it is already a model at mesoscopic scale.
Neurons themselves have significant inhomogeneity -- even neurons of same
cell type show significant variation between one another.
The coefficients $\An$ are supposed to model \textit{on-site} property of
a group of neurons hence they are likely to vary in space in a random
manner.
As this emerges after averaging over a group of neurons we expect
this variation to be less than the variation between properties of
individual neurons.
We do not consider this variation in the present model. 

We note that the connection between groups of neurons is
also not a constant.
In particular, a group of neurons can have stronger connections to
another group in a particular direction than other directions.
The connections between groups of neurons are not necessarily
isotropic. We focus on modeling this random anisotropy.
 As the properties of the neurons do not change over the time scales
we consider~\footnote{In practice, neurons are plastic, i.e., their
  properties change with time but over a time scale that is
  much slower than the time scales we consider here. }
this noise to be quenched.
We assume that this noise is self-averaging the anisotropy averages to zero
at large scales -- it is a small scale quenched noise. 
This allows us to introduce a new noisy term to the model of 
Ref.~\citep{tiberi2022gell}, in particular, we consider the model:
\begin{subequations}
  \begin{align}
  &\delt \phi = (-\mm\cdot\grad + \nun\lap)\phi + I\/. \\
  \text{where}\quad &\avg{m_i(\xx)m_j(\yy)}\equiv M_{ij}=\delta_{ij}M(r)\/,\\
  \text{with}\quad &r=\mid \xx - \yy\mid\/,\\
  \text{and}\quad &\avg{I(\xx,t)I(\yy,s)} = \Dn\delta^d(r)\delta(t-s)\/\equiv
  \Dt.
\end{align}
\label{eq:modelB}
\end{subequations}
The correlator of the quenched noise is given by
\begin{equation}
  M(r) = \Mn f(r/a)
\end{equation}
where $\Mn$ is a constant and $f(r/a)$ is a function of $r$ with a
characteristic length scale $a$.
In other words, the local anisotropy is modeled by a vector noise $\mm$
which is Gaussian, zero mean, and covariance $M$. 
Here and henceforth we use the notation that repeated indices are summed.

\section{Results}
\subsection{Replica action}
To analyze the stochastic PDE given in \eq{eq:modelB}, we use the
MSRDJ (Martin-Siggia-Rose-De Dominicis-Janssen) path-integral
formalism ~\citep{martin10ha,janssen1976lagrangean,de1978field}.
In addition to the usual formalism, our model contains
quenched noise. 
Let us first consider this model with one realization of the
quenched noise.
We rewrite our model as
\begin{subequations}
  \begin{align}
    \mL\phi - I &= 0 \/, \\
    \text{where}\quad\mL &\equiv (\delt + \mm\cdot\grad - \nu\lap)\/.
  \end{align}
\end{subequations}
This is a stochastic partial differential equation.
The solution consists of finding out the space-time dependent
probability distribution function $\mP[\phi(\xx,t)]$.
In the MSRDJ formalism, we write down the corresponding moment
generating functional
\begin{equation}
  \mZ[\phi] = \int \mD\phi\mD I \delta(\mL\phi - I)\exp\left(-\half
                I\Dot\Dt^{-1}\Dot I\right)\/.
  \label{eq:mZ1}
\end{equation}
where $\Dt$ noise correlation on the right hand side of \eq{eq:noise}.
Here the generating functional is written for one realization of the
quenched noise $m$.
The symbol $\Dot$ is defined in the following way.
For two functions $f(\xx,t)$ and $g(\xx,t)$ and an operator $\mM$,
\begin{subequations}
  \begin{align}
  f\Dot\mM\Dot g &\equiv \int d^dxd^dydtds f(\xx,t)\mM(\xx,t,\yy,s)g(\yy,s) \\
  \text{and}\quad   f\Dot g &\equiv \int d^dxdt f(\xx,t)g(\xx,t)\/.
  \end{align}
\end{subequations}
Now we introduce an additional auxiliary field $\Phi$ to rewrite
the functional $\delta$ function in \eq{eq:mZ1} as
\begin{equation}
  \mZ[\phi] = \int \mD\Phi\mD\phi\mD I \exp
  \left[i\Phi\Dot(\mL\phi - I) -\half I\Dot\Dt^{-1}\Dot I\right].
  \label{eq:mZ2}
\end{equation}
Integrating over the Gaussian noise $I$, we obtain
\begin{equation}
  \mZ(J,B) = \int \mD\Phi\mD\phi \exp\left[i\Phi\Dot\mL\phi
    -\half \Phi\Dot\Dt\Dot\Phi  + J\Dot\phi + B\Dot\Phi\right].
  \label{eq:mZ3}
\end{equation}
Here, in addition, we have introduced two source functions $J(\xx,t)$ and
$B(\xx,t)$ such that taking functional derivatives with respect to them we
can calculate any moment of $\phi$ and $\Phi$. 
The path--integral is bilinear in $\phi$ and $\Phi$, i.e., it can
be evaluated exactly and all moments, e.g.,
$\avg{\Phi\phi}$ and $\avg{\phi\phi}$ can be calculated exactly. 
Then these moments must be averaged over the statistics of the
quenched noise $m$. 
If we are to calculate all the moments, then it is best
to calculate $\avgm{\ln \mZ(J,B)}$ where $\avgm{\cdot}$ denotes
averaging over the statistics of $m$.
The standard technique is to use the replica
trick~\citep{parisi1980order, Mezard1987spin}, which
starts by recognizing that
\begin{equation}
  \ln \mZ = \lim_{N\to 0} \frac{\mZ^N-1}{N}.
  \label{eq:replica}
\end{equation}
The trick consists of first calculating $\avgm{\mZ^N}$, for any integer $N$
and then taking the limit $N\to 0$.
The product of $N$ path-integrals is
\begin{widetext}
\begin{subequations}
  \begin{align}
    \mZ^N &= \int \prod_{\alpha}^{N}\mD\Phia\mD\phia
    \exp\left[\sum^{N}_{\alpha}\left(S^0_{\alpha}
         + i\Phia\Dot \mm\cdot\grad \phia\right)\right]. \\
    \text{where}\quad S^0_{\alpha} &\equiv i\Phia\Dot L \Dot\phia
    - \half\Phia\Dot\Dt\Dot\Phia
    \label{eq:Sa}\\
    \text{and}\quad L &\equiv \delt - \nu\lap.
  \end{align}
\end{subequations}
Next we average this product over the statistics of $m$:
\begin{subequations}
  \begin{align}
    \avgm{\mZ^N} &= \int \prod_{\alpha}^{N}\mD\Phia\mD\phia
    \exp\left[\sum^{N}_{\alpha}S_{\alpha}\right]
	  \int \mD m \exp\left[-\half m_i\Dot (M^{-1})_{ij}\Dot m_j +
             \sum^{N}_{\alpha}i\Phia\Dot \mm\cdot\grad\phia\right] \\
    &= \int \prod_{\alpha}^{N}\mD\Phia\mD\phia e^{\Srep}, \\
    \text{where}\quad
     \Srep &\equiv  \sum^{N}_{\alpha}S^0_{\alpha}-\half
	    \sum^{N}_{\alpha,\beta}\Phia\Dot\grad_i\phia\Dot M_{ij} \Dot\Phib\Dot\grad_j\phib.
  \end{align}
\label{eq:repZ}
\end{subequations}
\end{widetext}
Here in the last step we have done the Gaussian integration over the
distribution of $m$ to obtain the replica action, $\Srep$. 
The result is a  bi-quadratic term containing both $\Phi$ and $\phi$ coupling
the replicas together.
This introduces new effective nonlinearity in our model due to averaging
over the noise.
The strength of this nonlinear is proportional to $\Mn$. 
\subsection{Renormalization group analysis}
Henceforth we follow the standard prescription of Wilsonian momentum shell
renormalization
group~\citep[see, e.g.,][for a pedagogical introduction]{shankar2017quantum}.
Details of the calculation are given in Appendix~\ref{sec:RGB}.
The calculations are done in Fourier space:
\begin{subequations}
  \begin{align}
    \phi(\kk,\omega) \equiv
     \int \phi(\xx,t)e^{i\kk\cdot\xx-i\omega t}d^dx dt, \\
     \Phi(\kk,\omega) \equiv \int \Phi(\xx,t)
                 e^{i\kk\cdot\xx-i\omega t}d^dx dt.
  \end{align}
\end{subequations}
To avoid proliferation of symbols, we use the same symbol for a field
in real and Fourier space.
In case of possible confusion, we distinguish them by explicitly giving
their argument. 
In Fourier space our problem has a high $k$ cutoff (ultraviolet cutoff)
$\Lambda$.
We consider a thin shell in Fourier space between $\Lambda/b$ to $\Lambda$.
Later we shall take the limit $b\to 1$.
We separate both $\phi$ and $\Phi$ in two, one with wavevector
$\mid\kk\mid$ less than $\Lambda/b$ and the other with wavevector
lying within the thin shell $\Lambda/b$ to $\Lambda$.
They are denoted respectively by $\phil$ ($\Phil$) and
$\phig$($\Phig$), i.e.,
\begin{subequations}
  \begin{align}
    \phi &= \phil + \phig, \\
    \Phi &= \Phil + \Phig. 
  \end{align}
\end{subequations}
The fields with label $>$ are called ``fast'' modes
and with label $<$ are called ``slow''.
As we eventually take the limit $b \to 1$, we rewrite
$b = \exp(\dl) \approx 1 + \dl$.
The key idea of Wilsonian RG is to integrate over the fast modes to write
an effective theory for the slow mode.
The crucial limitation is that the effective
theory is constrained to have the same functional form as
the original one we started with but with coupling constants --
$\Dn$, $\nun$ and $\Mn$ -- each becoming a function of scale, $\ell$.

In the limit $N \rightarrow 0$ and at the level of one-loop,
the RG flow equations for $D$, $\nu$, and $M$  are:
\begin{subequations}
\begin{align}
    \frac{d \nu}{d\ell} &= \nu \left[z-2 +
    \frac{g}{2}
      (1-\frac{2}{d}) \right]= \beta_{\rm \nu} \\
      \frac{d M}{d\ell} &= M \left(2z  -2 \right)= \beta_{\rm M} \\
    \frac{d D}{d\ell} &= D \left( z-d-2\alpha + g  \right)= \beta_{\rm D}
\end{align}
\end{subequations}
where we have set $\Lambda=1$ and defined
$g\equiv(M\fh(a)\Kd/\nu^2)$ as the effective coupling constant.
The righ hand side of the RG flow equations are called the $\beta$-functions.
\begin{figure}
  \includegraphics[width=\columnwidth]{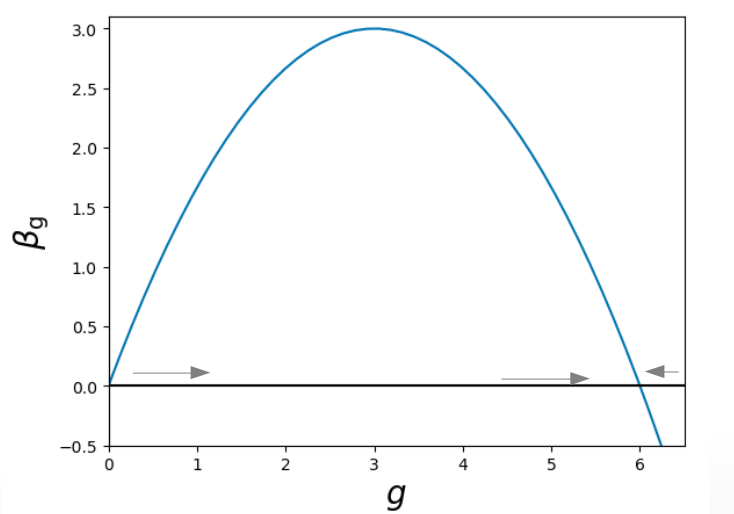}
  \caption{\label{fig:gbeta} The function $\beta_{\rm g}$
    as a function of the effective nonlinear coupling $g$ for $d=3$.
    The two fixed points are at zero and $6$, where the
    function $\beta_{\rm g}$ cuts the abscissa. The arrows
    show the RG flow. The fixed point at $0$ is unstable
  and the one at $6$ is stable. }
\end{figure}
\subsubsection{Fixed points and critical exponents}
Thus we obtain the RG flow equaion for $g$:
\begin{equation}
  \frac{dg}{d\ell} =g\left[ 2 - g\left\{1 - \frac{2}{d} \right\}\right]
  = \beta_{\rm g}
  \label{eq:dgdl}
\end{equation}
There are two fixed points, $\gastn=0$ and
\begin{equation}
  \gasto = \frac{2d}{d-2}
\label{eq:gasto}
\end{equation}
At the fixed points all the $\beta$-functions must be zero. 
At $g=\gastn = 0 $, $M=0$, 
\begin{equation}
  z = 2\/, \quad\text{and}\quad \alpha = 1-d/2
  \label{eq:fp1}
\end{equation}
At $g=\gasto$
\begin{equation}
  z = 1\/, \quad\text{and}\quad \alpha = \frac{z-d}{2} + \frac{d}{d-2}
  \label{eq:fp1}
\end{equation}
For $d>\dc = 2$ the fixed point $\gasto$ is positive and stable
where the fixed point $\gastn$ is unstable.
This is sketched in \Fig{fig:gbeta} for $d=3$ .
At exactly $d=2$ the fixed point $\gasto$ is at infinity.
It is possible that similar to KPZ this singularity is an artifact
of the one loop calculation~\citep{frey1994two}. 
For $d<\dc$ the fixed point $\gasto$ gives unphysical values.
This is not accessible within the perturbation theory.

Thus we reach the key conclusion that for $d>2$
the effect of the quenched noise is to change the
large scale dynamic behavior from diffusive ($z=2$) to
advective ($z=1$). 

\section{Summary}
Our goal in this paper is to introduce the framework
of replica renormalization group to problems in neuroscience.
The replica method has been successfully applied to
a large class of problems with quenched noise
in equilibrium statistical mechanics~\citep{Mezard1987spin}.
Given the natural heterogeneity of neural networks
in brain it seems to be well suited to extract
large scale behavior in neuroscience too.
We apply it to a model in which the connectivity of the neurons is
anisotropic.
Our calculations show that the presence of quenched noise can
fundamentally change the dynamic behavior of the system
from diffusive to advective. 
Although it remains to be seen whether this result
remains valid at higher orders in perturbation theory.
\begin{acknowledgments}
We thank Hauke Wernecke for providing the code to simulate Perlin noise 
(Fig. 1). 
This work was funded in parts by StratNeuro (to AK), 
Digital Futures grants (to AK and PH), the Inst. of Advanced Studies, 
University of Strasbourg, France Fellowship (to AK), 
and the National Natural Science Foundation of China under Grant No.
11572127 and 11872183 (to SL).
DM acknowledges the support of the Swedish Research Council Grant 
No. 638-2013-9243.
DM thanks Rahul Pandit for useful discussions. 
\end{acknowledgments}
\bibliography{neuro}%
\appendix
\section{Replica Renormalization group analysis}
\label{sec:RGB}
\begin{figure}
  \includegraphics[width=0.8\columnwidth]{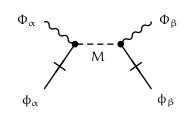}
  \caption{\label{fig:vertex}The Feynman graph for the interaction part of the
    action in \eq{eq:SI}.}
\end{figure}

The model is given by
\begin{subequations}
  \begin{align}
  &\delt \phi = (-\mm\cdot\grad + \nun\lap)\phi + I\/. \\
  \text{where}\quad &\avg{m_i(\xx)m_j(\yy)}\equiv M_{ij}=\delta_{ij}M(r)\/,\\
  \text{with}\quad &r=\mid \xx - \yy\mid\/,\\
  \text{and}\quad &\avg{I(\xx,t)I(\yy,s)} = \Dn\delta^d(r)\delta(t-s)\/\equiv \Dt.
\end{align}
\label{eq:modelB}
\end{subequations}
The correlator of the quenched noise is given by
\begin{equation}
  M(r) = \Mn f(r/a)
\end{equation}
where $\Mn$ is a constant and $f(r/a)$ is a function of $r$ with a
characteristic length scale $a$.
We assume that under RG the coupling constant $\Mn, \nun$ and $\Dn$ renormalizes
but the function $f$ (or $\fh$) remains unchanged.
To write the model is Fourier space, we define
\begin{subequations}
  \begin{align}
    \phi(\kk,\omega) &= \int\phi(\xx,t)\exp i(\qq\cdot\xx-\omega t) d^dxdt \\
    \phi(\xx,t) &= \int\phi(\kk,\omega)\exp -i(\qq\cdot\xx-\omega t)
    \frac{d^dkd\omega}{(2\pi)^{d+1}}
  \end{align}
\end{subequations}
The quenched noise correlation in Fourier space takes the form:
\begin{subequations}
  \begin{align}
    \avg{m_i(\pp)m_j(\qq)} &=\Mn\delta_{ij}\delta(\pp+\qq)\fh(aq) \\
    \text{where}\quad \fh(aq) &= \int f\left(\frac{r}{a}\right)
    e^{-i\qq\cdot\rr}d^dr
  \end{align}
\end{subequations}
The action corresponding to this model is
\begin{widetext}
\begin{subequations}
  \begin{align}
    \Srep &= S^{\rm 0}+\SI \\
    S^{\rm 0} &= \int_{\kk,\omega}\Phi(\kv)\left[\frac{\Dn}{2}\Phi(-\kv)
          -(i\omega+\nu k^2)\phi(-\kv)\right] \\
\SI &=
\frac{i^2\Mn}{2} \sum^{N}_{\alpha,\beta}\int_{\left\{\kv_{\rm i},\pp,\qq\right\}}
(\kkt\cdot\kkf)\fh(ap)
\Phia(\kvo)\phia(\kvt)\Phib(\kvth)\phib(\kvf) \\
&\times\delta(\kko+\kkt+\pp)\delta(\pp+\qq)\delta(\kkth+\kkf+\qq)
\delta(\oo+\ot)\delta(\oth+\of)
\label{eq:SI}
\end{align}
\end{subequations}
\end{widetext}
The Feynman graph of the interacting part of the action is shown in
\Fig{fig:vertex}.
Here we connect $\phia(\kvt)$ and $\Phib(\kvth)$, then we have
\begin{subequations}
\begin{align}
    \kkth+\kkt& = 0\\
    \kko + \kkt +\pp & = 0\\
    \kkth + \kkf -\pp & = 0  
\end{align}
\end{subequations}
which means
\begin{subequations}
\begin{align}
    \kkt& = -\pp - \kko\\
    \kkth  & = -\kkt = \pp + \kko\\
    \kkf  & = \pp - \kkth = -\kko.
\end{align}
\end{subequations}
Then we have
\begin{subequations}
\begin{align}
    \kkt \cdot \kkf &= \kko \cdot (\pp+\kko) \\
    \avg{\phia(\kvt)\Phib(\kvth)}_{0} &\to   \dab G_0(-\pp-\kko,\omega)
\end{align}
\end{subequations}
Here the symbol $\kk$ denotes a vector in $d$ dimensional space and
the symbol $\kv$ denotes $(\kk,\omega)$. 
To obtain the ``naive'' dimensions we use the fact that 
the action must be dimensionless and use $\Lambda$ and $\nun\Lambda^2$
are our unit of one-over-length and frequencey respectively.
In other words:
$[q] = \Lambda$ and $[\omega] = \nun\Lambda^2$,
and $[\fh] = \Lambda^{-d}$
We obtain
\begin{subequations}
  \begin{align}
    [\phi] &= \Lambda^{-d/2-3}\nun^{-3/2}\Dn^{1/2} \\
    [\Phi] &= \Lambda^{-d/2-1}\nun^{-1/2}\Dn^{-1/2}\\
    [\Mn] &= \Lambda^{2-2d}\nun^2
  \end{align}
\end{subequations}
The effective coupling constant is $g = M/\nu^2$ whose 
engineering is $[\Mn/\nun^2] = \Lambda^{2-2d}$.
The coupling constant is dimensionless at $d=1$.

The ``bare'' Green's function and the ``bare'' correlation functions,
which are the same as the KPZ problem~\citep{kardar1986dynamic,
  forster1977large, medina1989burgers, frey1994two}, are respectively:
\begin{subequations}
  \begin{align}
    G_0(\kk,\omega) &= \frac{1}{\nun k^2 - i\omega} \\
    C_0(\kk,\omega) &= \frac{\Dn}{(\nun k^2 - i\omega)(\nun k^2 + i\omega)}
  \end{align}
  \label{eq:free}
\end{subequations}
The RG flow equations are obtained by two steps; 
decimation and  rescaling. 
\subsubsection{Decimation}
\begin{figure*}[t!]
\centering
\includegraphics[width=\textwidth]{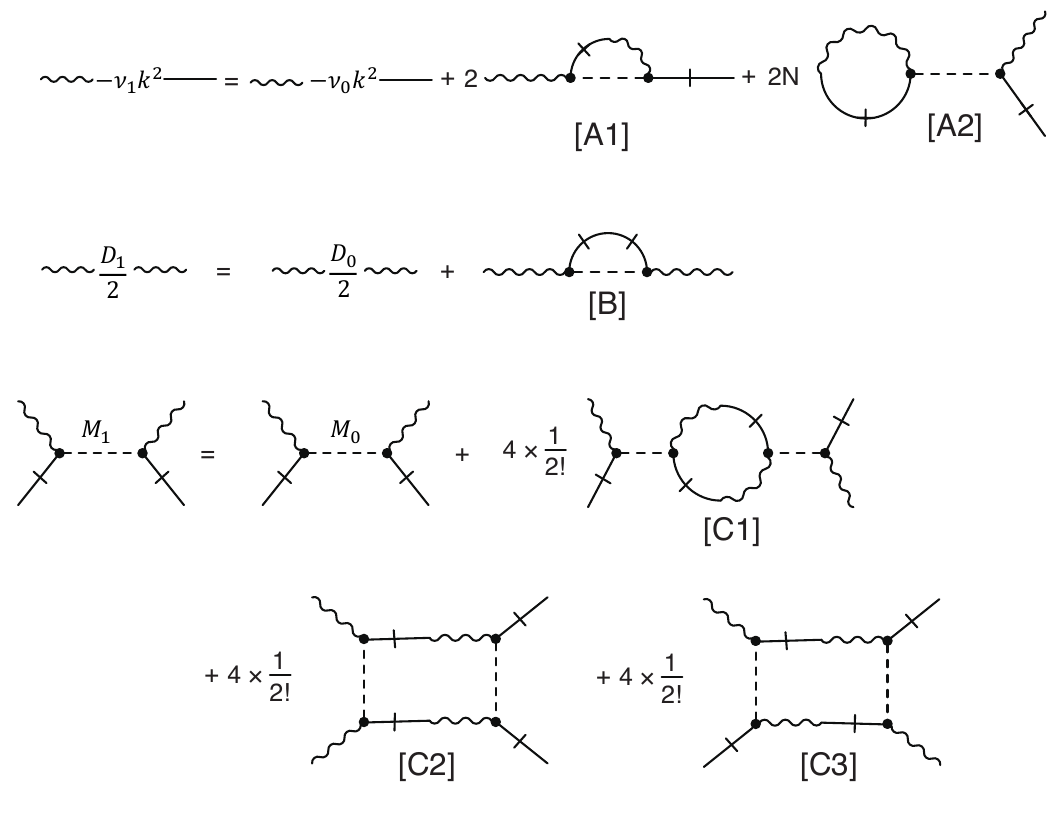}
\caption{The Feynman diagrams to calculate correction to $\nu$, $D$ and
  $M$ up to one loop order. Short straight lines denote $\phi$, short wavy
  lines denote $\Phi$, long straight lines denote the
  free correlation function 
  and long mixed lines (one end wavy, one end straight) denote free Green's
  function. Dashed lines denote the noise correlation.
  A cut on a line shows a derivative (in real space) or
  multiplication by a wavenumber in Fourier space.  
}\label{fig:diagB}
\end{figure*}

We integrate over the high $k$ modes over a thin shell in Fourier space,
$\Lambda/b$ to $\Lambda$ where $\Lambda$ is the high $k$ cutoff of
our model, 
and absorb the result by renormalizing the coupling constant
while keeping the functional form of the action the same.
It is convenient to do this integration perturbatively
using Feynman diagrams. 
The Feynman diagrams to one loop order are shown in \Fig{fig:diagB}.

As shown in \Fig{fig:diagB}, the corrections to  $\nun$, $\Dn$, and
$\Mn$ are as follows:
\begin{subequations}
    \begin{align}
        A1 &=  -\frac{\Mn \fh(a\Lambda)\Lambda^{d-2}\Kd}{4 \nun}  \dl
     \left(1 - \frac{2}{d}\right)k^2  \\
        A2 &=  0  \\
        B &=  \frac{\Mn \Dn \fh(a\Lambda)\Lambda^{d-2}\Kd}{2\nun^2}  \dl  \\
         C1 & =   0  \\
        C2 & =  - C3  
    \end{align}
\end{subequations}
The detail of the integrals are given in section~\ref{sec:FIntB}.
At one loop order the renormalized coupling constants are
\begin{subequations}
  \begin{align}
    \nuo &= \nun\left[ 1 + \frac{\Mn\fh(a \Lambda)\Lambda^{d-2}\Kd}{2 \nun^2}
      \left(1-\frac{2}{d}\right)\dl \right] \\
    \Mo &= \Mn\\
    \Do &= \Dn\left[1 + \frac{\Mn\fh(a\Lambda)\Lambda^{d-2}\Kd}{ \nun^2}\dl \right]
  \end{align}
  \label{eq:one}
\end{subequations}
where we have defined $P = \fh(a\Lambda)$ and $\xi = \fh(a\Lambda)/\fh(0)$. 
\subsubsection{Rescaling}
Our model now has a new cutoff $\Lambda/b$.
We rescale to obtain a model as close to the original one as possible.
Under rescaling by $x\to bx$, $k \to k/b$, $t \to b^z t$,
$\phi \to b^{\alpha}\phi$ and demanding that the governing equations
remain unchanged  we obtain
\begin{subequations}
  \begin{align}
    \nun &\to b^{z-2}\nun \\
    \Mn &\to b^{2z-2}\Mn \\
    \Dn &\to b^{z-d-2\alpha}\Dn
  \end{align}
\end{subequations}
As a result of decimation and rescaling we now obtain the
RG flow equations:
\begin{subequations}
\begin{align}
    \frac{d \nu}{d\ell} &= \nu \left[z-2 +
    \frac{g}{2}
      (1-\frac{2}{d}) \right]= \beta_{\rm \nu} \\
      \frac{d M}{d\ell} &= M \left(2z  -2 \right)= \beta_{\rm M} \\
    \frac{d D}{d\ell} &= D \left( z-d-2\alpha + g  \right)= \beta_{\rm D}
\end{align}
\end{subequations}
where we have set $\Lambda=1$ and defined
$g\equiv(M\fh(a)\Kd/\nu^2)$ as the effective coupling constant.
The righ hand side of the RG flow equations are called the $\beta$-functions.

\subsection{Integrals appearing in the Feynman diagrams}
\label{sec:FIntB}
In what follows, we first non-dimensionalize the integrals by
using $[k] = \Lambda$ for wavenumber and $[\omega] = \nun\Lambda^2$
for frequency. But we use the same symbols for the non-dimensional
quantities. More specifically
\begin{subequations}
  \begin{align}
    \intg\frac{d^dk}{(2\pi)^d} &\to
              \Lambda^d \int_{1/b}^{1}\frac{d^dk}{(2\pi)^d} \\
    \int_{-\infty}^{\infty}\frac{d\omega}{(2\pi)} &\to
               \nun\Lambda^2 \int_{-\infty}^{\infty}\frac{d\omega}{(2\pi)} \\
    G_0(\kk,\omega) &\to \frac{1}{\nun\Lambda^2}\frac{1}{k^2-i\omega} \\
    C_0(\kk,\omega) &\to \frac{D_0}{\nun^2\Lambda^4}
    \frac{1}{(k^2-i\omega)(k^2+i\omega)} 
  \end{align}
  \end{subequations}

In evaluating the integrals over the inner momentum variable it is useful to
define
$\kkp = \kk/2+\pp$ and $\kkm = \kk/2-\pp$.
Typically, all the integrals are functions of $\kk$ and $\omega$.
We take the limit of $\omega \to 0$ and $k \to 0$ and keep the
leading order term.
The integral of any function $F(p)$ over the interval
$1/b$ to $1$ is
\begin{equation}
  \int_{1/b}^{1}F(p)dp = F(1)\dl
\end{equation}
where we have used $b=\exp(\dl)$ and taken the limit $\dl\to 0$. 
The integrals corresponding to the diagrams are as follows:
\begin{widetext}
  \begin{subequations}
    \begin{align}
      A &= -\frac{i^2\Mn}{2}\intg_{\pp}G_0(\kkm,\omega)\fh(a\kp)(\kkm\cdot\kk) \\
      &= -\frac{\Mn\Lambda^{2d}}{2\nun}\intg_{\pp}
      \frac{\fh(a\kp)(\kkm\cdot\kk)}{\km^2-i\omega} \\
      &= -\frac{\Mn\Lambda^{2d}}{2\nun}
      \intg_{\pp}\frac{\fh(a\kp)}{p^2}\left(-kp\cos(\theta)+
      \frac{k^2}{2} -k^2\cos^2(\theta) + \hot \right)
      \label{eq:A3}\\
      &\approx -\frac{\Mn\Lambda^{2d}\Kd}{4\nun}
      \left(1-\frac{2}{d}\right)\fh(a\Lambda)k^2\dl
    \end{align}
  \end{subequations}

 The angular integral in the first term within the parenthesis
  in \eq{eq:A3} is zero.
  Hence in the limit $k\to 0$ the leading order contribution in \eq{eq:A3} in
  proportional to $k^2$.
  Note that integral is over a shell in Fourier space between
  $p = \Lambda/b$ to $\Lambda$.
  It is reasonable to assume that the function $\fh(\xi)$ goes to a
  constant in the limit $\xi \to 0$. 
  If we had extended the integral from $p=0$ to $\Lambda$ the integral
  would have blown up in the lower limit.
  For this reason a naive perturbation theory would have failed. 
  Similar infra-red divergence that also appears in
  naive perturbation theory of KPZ equation~\citep{frey1994two}.
  Like the KPZ problem, DRG is able to control this divergence
  in the present problem. 

  \begin{subequations}
    \begin{align}
      B &= -\frac{i^2\Mn}{2}\intg_{\pp}C_0(\kkm,\omega)\fh(a\kp)(\kkm\cdot\kkm) \\
      &= \frac{\Mn\Dn\Lambda^{d-2}}{2\nun^2}
      \intg_{\pp}\frac{\fh(a\kp)\km^2}{\km^4} \\
      &= \frac{\Mn\Dn\Lambda^{d-2}}{2\nun^2}\intg\frac{\fh(a\kp)}{p^2}
      \frac{d^dp}{(2\pi)^{d}} +\hot\\
      &\approx \frac{\Mn\Dn\Lambda^{d-2}\Kd}{2\nun^2}\fh(a\Lambda)\dl
    \end{align}
  \end{subequations}

  \begin{subequations}
    \begin{align}
      C2 &= -\left(\frac{i^2\Mn}{2}\right)^2\intg_{\pp,\sigma}
      G_0(\kkp,\sigma)G_0(-\kkm,-\sigma)(\kkm\cdot\kkp) \\
      &=\frac{\Mn^2\Lambda^{d-2}}{4\nun}
      \intg_{\pp}\int \frac{d\sigma}{2\pi}
      \frac{(\kkm\cdot\kkp)}{(\sigma+i\kp^2)(\sigma+i\km^2)} =0
      \label{eq:C22}
    \end{align}
  \end{subequations}
  In \eq{eq:C22} the integral of $\sigma$ is zero. 

\end{widetext}
\section{Useful identities in d dimensions}
The surface area and volume of the unit sphere in $d$ dimensions
are, respectively,
\begin{subequations}
  \begin{align}
    \Sd &= \frac{2\pi^{d/2}}{\Gamma(d/2)}\\
    \Vd &= \frac{\pi^{d/2}}{\Gamma(d/2 + 1)}
    \end{align}
  \label{eq:Sd}
\end{subequations}
where $\Gamma$ is the Gamma function. 
Volume element in spherical coordinate in $d$ dimensions
\begin{equation}
  d^dV = r^{d-1}\sin^{d-2}(\theta_{\rm 1})\sin^{d-3}(\theta_{\rm 2})\ldots \sin^{d-2}(\theta_{\rm d-2}) drd\theta_{\rm 1}\ldots d\theta_{\rm d-1}
\end{equation}
The integration of a function that depends only on the magnitude of
wavevector $\kk$ in $d$ dimensions
\begin{equation}
  \int f(k)d^dk = \Sd\int_0^{\infty}f(k)k^{d-1}dk
\end{equation}
Few other useful integrals~\citep[see, e.g.,][Appendix B]{Barabasi1995fractal}
\begin{subequations}
  \begin{align}
    \int_0^{\pi}\sin^{d-2}\theta d\theta &= \frac{\Sd}{S_{\rm d-1}} \\
    \int_0^{\pi}\sin^{d-2}\theta\cos\theta &= 0 \\
    \int_0^{\pi}\sin^{d-2}\theta\cos^2\theta &= \frac{1}{d}\frac{\Sd}{S_{\rm d-1}} 
  \end{align}
\end{subequations}


\end{document}